\documentclass[epj]{svjour}
\usepackage{graphics}
\begin{document}
\title{Electrostatically induced undulations of lamellar DNA--lipid complexes}
\author{Helmut Schiessel\inst{1} \and Helim Aranda-Espinoza\inst{2}}
%
\institute{Max-Planck-Institute for Polymer Research, Theory
Group, POBox 3148, D-55021 Mainz, Germany \and Institute for
Medicine and Engineering, University of Pennsylvania,
Philadelphia, PA 19104, USA}
\date{Received: date / Revised version: date}
%
\abstract{ We consider DNA-cationic lipid complexes that form
lamellar stacks of lipid bilayers with parallel DNA strands
intercalated in between. We calculate the electrostatically
induced elastic deformations of the lipid bilayers. It is found
that the membranes undulate with a periodicity that is set by the
DNA interaxial distance. As a consequence the lamellar repeat
distance changes resulting in a swelling or compression of the
lamellar stack. Such undulations may be responsible for the
intermembrane coupling between DNA strands in different layers as
it is observed experimentally.
} 
\maketitle

\section{Introduction}

\label{intro} Electrostatic adsorption of polyelectrolytes onto oppositely
charged surfaces, such as lipid membranes, has been the subject of intense
experimental and theoretical research in the last decade. Of particular
interest is the spontaneous complexation of DNA with both cationic and
neutral lipids due to their possible application to gene therapy \cite
{Felgner97,Crystal95}. These so-called ''lipoplexes'' show a diversity of
equilibrium and metastable structures \cite
{Sternberg94,lasic97,Raedler97,fang97,salditt97,koltover98,Artzner98,Zantl98,huebner99,Koltover99}%
. For example, it has been shown through X-ray diffraction analysis \cite
{Raedler97,salditt97,Koltover99} that DNA molecules and lipids can form
lamellar complexes with DNA intercalated in between lipid bilayers. Another
complex formed for sufficiently flexible membranes is an inverse hexagonal
lipid structure with DNA inside the water regions \cite{koltover98}.

Several theoretical studies help to understand many of the phenomena
observed for lipoplexes \cite
{dan96,dan97,bruinsma98b,May97,dan98,bruinsma98,golubovic98,ohern98b,Harries98,Schiessel98,menes98,ohern98,ohern99,may00,menes00}%
, however many more remain to be elucidated. May et al. \cite{may00} studied
in detail the phase behavior of aqueous mixtures of DNA, cationic lipid and
neutral lipid. Their model is based on the two thermodynamically stable
structures found experimentally \cite{koltover98} and predicts the phase
diagram as a function of the DNA/lipid composition and the elastic
properties of the lipid bilayers. Another problem that was considered
theoretically is the dependence of the interaxial spacing of DNA rods in
lamellar complexes on the DNA/lipid composition. Bruinsma \cite{bruinsma98}
presented an analytical approach that is applicable to lipoplexes with
weakly charged bilayers. The numerical study of Harries \textit{et al.} \cite
{Harries98} predicts the interaxial spacing also for higher charge
densities. According to both studies the isoelectric point of the lipoplex
(the point at which the anionic charges of the DNA balance the cationic
charges of the lipids) is unstable to further adsorption of DNA or lipids.
The formation of say an isoelectric complex is driven by the release of the
small counterions that were ''condensed'' on the highly charged DNA and on
the charged bilayer before complexation. A lipoplex close to the isoelectric
point is very susceptible to the uptake of further cationic lipids or DNA --
if available -- since this will be accompanied by the release of the
corresponding counterions into the lipoplex\footnote{%
A similar instability is also expected for the complexation of charged
spheres and oppositely charged polyelectrolytes. A single highly charged
chain will wrap around a single sphere forming a complex that is beyond the
isoelectric point (''overcharging'') and this effect is driven by the
release of counterions from the wrapped chain\cite{park99}. On the other
hand, a chain in a solution of highly charged spheres will complex more
spheres than necessary to be isoelectric, and this effect is driven by the
release of counterions of the complexed spheres\cite{Schiessel00}.}. The
theoretical predictions show good agreement with a recent experimental study
\cite{Koltover99}.

A different approach to the problem of the distance between DNA strands was
given by Dan \cite{dan96}. In this study the preferred distance was
predicted to be the result of two competitve mechanisms, electrostatic
repulsion between the strands and their membrane-induced attraction due to
the perturbation of the lipid packing in the membrane close to the adsorbed
DNA. The model assumes DNA strands adsorbed on a single membrane as it was
investigated experimentally by Fang and Yang \cite{fang97} using atomic
force microscopy experiments. They found that the distance between DNA
strands was about 5 nm, a distance that was predicted by Dan within her
model \cite{dan96}.

In that model it is assumed that the membrane (that is supported by a solid
surface) is locally perturbed close to the DNA in such a way that the
monolayer thickness is slightly increased \cite{dan96}. It should be
expected that for lamellar lipoplexes one also has perturbations. Since the
membranes are allowed to undergo shape changes freely (no supporting layer)
one might expect undulations leading to a compression or sweeling of the
whole lamellar stack as depicted in Fig. 2. Such undulations might lead to
an intermembrane coupling between DNA rods in different layers -- resulting
in a 3D ordering of the DNA rods. Lipoplexes with a 3D rectangular ordering
of the DNA molecules were indeed observed experimentally \cite{Artzner98}.

Most experiments on lamellar lipoplexes indicate that such a type of
perturbation of the membranes around the DNA molecules -- if present -- is
small\cite{Raedler97,salditt97}. Undulations of the membranes should lead to
a lamellar repeat distance that is larger or smaller than the sum of the
bilayer thickness and the diameter of the DNA molecule (including a
hydration shell). Considering complexes at the isoelectric point and
changing the ratio of charged to neutral lipids it was observed that the
lamellar repeat distance stays always close to a value that indicates flat
membranes\cite{Raedler97,salditt97}. Thus even though the lipid dilution
experiments lead to a considerable increase of the interaxial spacing
between DNA rods, the undulations remain too small to be non-ambiguously
detected. On the other hand, for more flexible membranes where detectable
membrane undulations could be expected the system switches to the inverse
hexagonal phase instead \cite{koltover98}.

Recently Subramanian \textit{et al. }\cite{Subu00} studied the complexation
of the anionic polypeptide poly-glutamic acid with a mixture of cationic
(DDAB) and neutral (DLPC) lipids by means of small angle X-ray scattering
and neutron scattering. It was observed that the lipid organizes in a
multilamellar phase with the polypeptide chains intercalated in between the
membranes. Compared to the DNA complexes discussed above, the polypeptides
do not show any in-plane ordering even though it is assumed that they are in
the $\alpha $-helical state. As for the DNA lipoplexes a ''lipid dilution''
experiment was performed for isoelectric polypeptide lipoplexes. Contrary to
the outcome for the DNA complexes, a considerable increase of the lamellar
spacing was found when the cationic lipids were diluted by neutral ones. For
high lipid dilution the spacing saturated at a constant value of 60\AA\
which coincides with the equilibrium value of pure DLPC membranes.
Subramanian \textit{et al.} \cite{Subu00} suggested that this behavior could
be due to a ''pinching mechanism'' including membrane undulations similar to
the ones depicted in Fig. 2 (case $h<0$). The pinching sites are formed due
to the electrostatic interaction between the negatively charged
poly-glutamic acid and the cationic DDAB lipids. Away from the pinched
regions the properties of the lipoplex are dominated by the properties of
the pure DLPC membranes. Whether it is possible to have pinches in a
lipoplex was studied by one of the authors \cite{Schiessel98}. By comparing
the gain in electrostatic free energy with the bending energy of forming a
pinch, the parameter range was estimated at which pinching can be expected.
It was shown that this effect should occur if the line charge density of the
rods is sufficiently high and the membranes are sufficiently flexible, a
situation that might be fulfilled for the polypeptide lipoplex considered in
Ref. \cite{Subu00}.

A different approach to the pinching problem is taken in the present study.
We start out with a perfectly flat lamellar lipoplex as depicted in Fig. 1.
The DNA rods are assumed to be ordered within a 3D rectangular lattice as it
was observed by Artzner \textit{et al.} \cite{Artzner98}. Our goal is to
calculate how the electrostatic interaction between the negatively charged
''rods'' and the positively charged membranes modifies the conformation of
the membranes. We show that there are in principle two possibilities, namely
a compression of the lamellar stack as depicted in Figure 2 ($h>0$) or an
expansion as depicted in the same Figure (case $h<0$).

In the next section we introduce the model system and calculate
its electrostatic and bending free energies for arbitrary but
small periodic undulations of the membranes. By minimizing the
free energies of the undulation with respect to its Fourier
components we show in Sect.~\ref{sec:2} that the electrostatic
interaction usually favors a compression of the lamellar complex
-- at least if the underlying assumptions of our model are
fulfilled. These assumptions are discussed in Sect.~\ref{sec:3}
where we also present some conclusions.

\section{Free energy of model lipoplex}

\label{sec:1} The aim of the following calculation is to determine
the electrostatic contribution to the undulations of a lamellar
stack of membranes with DNA molecules intercalated in between. Our
model system consists of two constituents, the membranes and the
DNA molecules. The membranes have a uniform thickness $t$ and
carry positive charges on both sides. The surface charge density
is given by $\sigma /2$ on each side of the bilayer and is assumed
to be uniform. In our model the membranes are perfectly
transparent for the electric field lines, i.e., we have a
homogeneous dielectric constant throughout the lipoplex. The
bilayers are flexible with a bending rigidity $k_{c}$. The DNA
molecules are modelled as infinitely long rigid rods of radius
$r$. For simplicity, we assume the negative charges of the DNA
molecules to be located along their middle axis with the linear
charge density $-\rho $. Following the experimental observation of
a lamellar stack with DNA forming smectic arrays we arrange the
components of our model in the following way (cf. Fig. 1). All
membranes are parallel to the $XY$-plane with their midplanes at
the positions $z=0,\pm 2\left( r+t/2\right) ,\pm 4\left(
r+t/2\right) ...$ The DNA rods are aligned in the $Y$-direction.
The interhelical spacing between neighboring DNA molecules is
constant and is denoted by $d=2\pi /q$. Excluded volume requires
that $d\geq 2r$. The rods in one layer are located at $x=0,\pm
d,\pm 2d...$, in the neighboring layers they are displaced by
$d/2$, i.e., they are at the positions $x=\pm d/2,\pm 3d/2,...$
etc. Furthermore, the rods are assumed to be always attached to
the two neighboring membranes.

\begin{figure}
\resizebox{0.48\textwidth}{!}{%
  \includegraphics{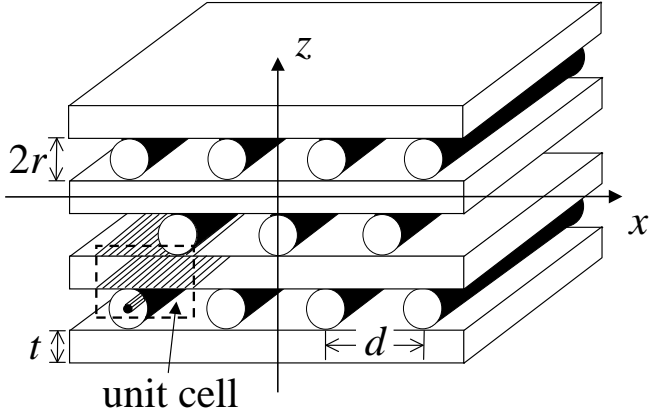}
}
\caption{Schematic view of the model lipoplex for the case of flat
membranes (see text for detail)}
\label{fig:1}       
\end{figure}

The electrostatic interaction between the charges is calculated
within the Debye-H\"{u}ckel approximation. In this approximation
the potential $\Phi $\ is determined by $\Delta \Phi =\kappa
^{2}\Phi $ with the appropriate boundary conditions. Here $\kappa
^{-1}$ denotes the Debye screening length
that is given by $\kappa ^{-1}=\left( 8\pi n_{s}l_{B}\right) ^{-1/2}$ where $%
n_{s}$ is the bulk salt concentration and $l_{B}=e^{2}/\varepsilon
k_{B}T$ is the Bjerrum length ($e$ is the unit charge, $k_{B}T$ is
the thermal energy and $\varepsilon $ the dielectric constant;
$\varepsilon \approx 80$ in an aqueous solution). The total
electrostatic contribution $F_{el}$ to the free energy of the
system is given by the sum of the (screened) electrostatic
interactions and the translational entropy of the counterions
\cite{Verwey,Goldstein90}:
\begin{equation}
F_{el}=\frac{1}{2}\int dS\sigma ^{\prime }\Phi  \label{fel}
\end{equation}
The integration extends over all charged surfaces of the system
with $\sigma ^{\prime }$ being the corresponding charge densities.

We ask the following question: How are the membranes deformed by
the electrostatic interaction? In order to answer this question we
will calculate the induced undulations of the membrane up to the
first order in the deformation amplitude.

Consider the membrane at $z=0$. Electrostatics induces a deformation $%
z=u\left( x\right) $ around the flat state, $z\equiv 0$. Due to
the symmetry the deformation profile is of the form
\begin{equation}
u\left( x\right) =\widehat{\sum_{n}}a_{n}\cos \left( nqx\right)
\label{prof}
\end{equation}
where the hat denotes summation over odd $n$ only. This undulation
leads to the following curvature energy $f_{bend}=\left(
k_{c}/2\right) \left( \nabla ^{2}u\right) ^{2}$ (per area $A$):
\begin{equation}
f_{bend}=\frac{k_{c}}{2}\widehat{\sum }_{n}a_{n}^{2}n^{4}q^{4}
\label{bend}
\end{equation}

In order to calculate the electrical free energy, Eq. \ref{fel},
we compute first the electrical potential $\Phi _{M}^{\left(
u\right) }\left( x,z\right) $ induced by the charges on the upper
surface of the membrane
(note that $\Phi _{M}^{\left( u\right) }$ is translational invariant in $Y$%
-direction). At that charged surface, i.e., at $z=t/2+u\left(
x\right) $, we have the boundary condition $\partial \Phi
_{M}^{\left( u\right) }/\partial n=-2\pi \sigma /\varepsilon $
which is here of the form
\begin{equation}
-\frac{\partial \Phi _{M}^{\left( u\right) }}{\partial x}\widehat{\sum_{n}}%
a_{n}nq\sin \left( nqx\right) +\frac{\partial \Phi _{M}^{\left( u\right) }}{%
\partial z}=-\frac{2\pi \sigma }{\varepsilon }  \label{bound}
\end{equation}
(up to terms of the order $a_{n}^{2}$). By expanding $\Phi
_{M}^{\left( u\right) }$ up to first order in the amplitudes
$a_{n}$ we find the
following form of the potential above the membrane ($z>t/2+u\left( x\right) $%
)
\begin{equation}
\Phi _{M}^{\left( u\right) }\left( x,z\right) =\varphi ^{\left(
0\right) }\left( x,z\right) +\widehat{\sum_{n}}a_{n}\varphi
^{\left( n\right) }\left( x,z\right)   \label{phim2}
\end{equation}
Each $\varphi ^{\left( n\right) }$ fulfills the Debye-H\"{u}ckel
equation separately. They can be expanded in Fourier series
$\varphi ^{\left( n\right) }=\sum_{m}B_{m}^{\left( n\right)
}\left( z\right) \\ \cos \left( mnqx\right) $ where $B_{m}^{\left(
n\right) }\left( z\right) =b_{m}^{\left(
n\right) }\exp \left( -\kappa _{nm}z\right) $ with $\kappa _{nm}=\sqrt{%
\kappa ^{2}+\left( nmq\right) ^{2}}$. The coefficients
$b_{m}^{\left( n\right) }$ follow from the boundary condition at
the membrane together with the fact that due to symmetry $\Phi
_{M}^{\left( u\right) }\left( x,z<t/2\right) \equiv \Phi
_{M}^{\left( u\right) }\left( \left( \pi /q\right) -x,t-z\right)
$. We find that only the coefficients $b_{1}^{\left( n\right) }$
are non-vanishing and are given by $b_{1}^{\left( n\right) }=\pi
\sigma \kappa /\varepsilon \kappa _{n}$. This leads to (for
$z>t/2$):
\begin{eqnarray}
\Phi _{M}^{\left( u\right) }\left( x,z\right)  &=&\frac{\pi \sigma }{%
\varepsilon \kappa }\left[ e^{-\kappa \left( z-t/2\right)
}+\right. \nonumber \\ &&\left.
\widehat{\sum_{n}}a_{n}\frac{\kappa ^{2}}{\kappa _{n}}e^{-\kappa
_{n}\left( z-t/2\right) }\cos \left( nqx\right) \right]
\label{phimu}
\end{eqnarray}
The total potential $\Phi _{M}$ induced by the membrane at
$z=u\left(
x\right) $ is the sum of the contributions of the upper charged boundary, $%
\Phi _{M}^{\left( u\right) }$, cf. Eq. \ref{phimu}, and of the lower one, $%
\Phi _{M}^{\left( l\right) }$: $\Phi _{M}=\Phi _{M}^{\left(
u\right) }+\Phi _{M}^{\left( l\right) }$. Using $\Phi _{M}^{\left(
l\right) }\left( x,z\right) =\Phi _{M}^{\left( u\right) }\left(
x,z+t\right) $ we find:
\begin{eqnarray}
\Phi _{M}\left( x,z\right)  &=&\frac{2\pi \sigma }{\varepsilon
\kappa }\left[ \cosh \left( \frac{\kappa t}{2}\right) e^{-\kappa
z}+\right.   \nonumber \\ &&\left.
\widehat{\sum_{n}}a_{n}\frac{\kappa ^{2}}{\kappa _{n}}\cosh \left(
\frac{\kappa _{n}t}{2}\right) e^{-\kappa _{n}z}\cos \left(
nqx\right) \right] \label{phim}
\end{eqnarray}
Furthermore, the potential induced by the line charge of the rod
has the form
\begin{equation}
\Phi _{R}\left( R\right) =-\frac{2\rho }{\varepsilon }K_{0}\left(
\kappa R\right)   \label{phir}
\end{equation}
where $R$ is the distance from the line and $K_{0}$ is a modified
Bessel
function with $K_{0}\left( x\right) \simeq -\ln x$ for $x\ll 1$ and $%
K_{0}\left( x\right) \simeq \left( \pi /2x\right) ^{1/2}\exp
\left( -x\right) $ for $x\gg 1$. The total electrical potential
$\Phi $ is the sum of the potential $\Phi _{1}$ that follows from
all membranes and the potential $\Phi _{2}$ that is due to all the
rods: $\Phi =\Phi _{1}+\Phi _{2} $.

We calculate now the total electrostatic contribution to the free
energy per unit cell. A unit cell has the width $d$ (in
$X$--direction) and a height that corresponds to the (average)
distance between neighboring layers (for the case of a flat
membrane -- depicted in Fig. 1 -- this height equals $2r+t $).
According to Eq. \ref{fel}, we obtain the total electrostatic
energy (per unit cell) by integrating the total potential over all
charged surfaces that lie within this cell. The unit cell in Fig.
1 contains three charged
surfaces, $S_{M}^{\left( u\right) }$, $S_{M}^{\left( l\right) }$ and $S_{R}$%
. $S_{M}^{\left( u\right) }$ is a stripe of the upper surface of
one bilayer that is uniformly charged with the density $\sigma
/2$. $S_{M}^{\left( l\right) }$ is the corresponding lower charged
surface. $S_{R}$ is the surface carrying the charges of one rod;
we assume this to be the surface of a cylinder with radius $\delta
r\ll r$ and charge density $-\sigma _{R}=-\rho /2\pi \delta r$. It
follows that the electrostatic energy (per
area) has three contributions: The inter-(and intra-) membrane interaction $%
f_{M}$, the membrane-rod interaction $f_{MR}$ and the interaction
between the rods $f_{R}$. Thus
\begin{eqnarray}
f_{el} = f_{M}+f_{MR}+f_{R} =\frac{\sigma
/2}{2A}\int_{S_{M}}dS\Phi _{1}- \nonumber \\
\frac{\sigma _{R}}{A}%
\int_{S_{R}}dS\Phi _{1}-\frac{\sigma _{R}}{2A}\int_{S_{R}}dS\Phi
_{2}
\end{eqnarray}
Here we made use of the identity $-\int_{S_{R}}dS\sigma _{R}\Phi
_{1} = \int_{S_{M}}dS\\ \frac{\sigma }{2}\Phi _{2}$.

\begin{figure}
\resizebox{0.48\textwidth}{!}{%
  \includegraphics{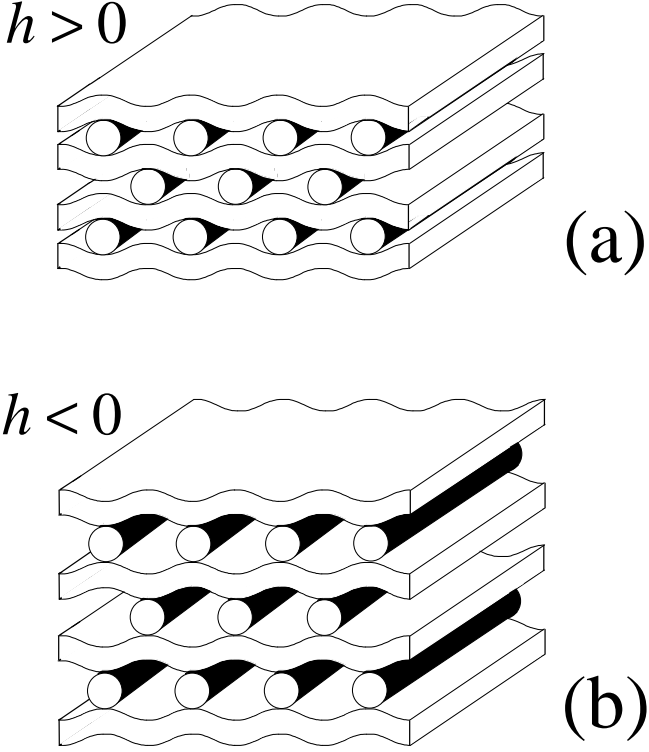}
}
\caption{Membrane undulations in lipoplexes. Shown are the two
cases (a) $h>0$: compression and (b) $h<0$: swelling of the
lamellar stack}
\label{fig:2}       
\end{figure}

We start by calculating the change of the membrane--membrane interactions $%
F_{M}$ induced by their undulations (up to first order in
$a_{n}$). The position of the midplane of the $k$th membrane is
given by
\begin{equation}
u_{k}\left( x\right) =2k\left( r+t/2-h\right) +\left( -1\right) ^{k}\widehat{%
\sum }_{n}a_{n}\cos \left( nqx\right)  \label{uk}
\end{equation}
with $k=0,\pm 1,\pm 2,...$ and $h=\widehat{\sum }_{n}a_{n}$. Fig.
2 shows
schematic views of lamellar structures that are compressed -- case (a) with $%
h>0$ -- and swollen -- case (b) with $h<0$. Denote the
contribution of $k$th membrane to the potential by $\Phi
_{M}^{\left( k\right) }$. Then

\begin{eqnarray}
&f_{M}& = \frac{q\sigma }{4\pi }\sum_{k=-\infty }^{\infty
}\int_{0}^{2\pi/q}dx\Phi _{M}^{\left( k\right) }\left( x,\frac{t}{2}+\widehat{\sum }%
_{n}a_{n}\cos \left( nqx\right) \right)   \nonumber \\
&\simeq &\frac{\pi \sigma ^{2}}{\varepsilon \kappa }\frac{\cosh \left( \frac{%
\kappa t}{2}\right) \cosh \left( \kappa r\right) }{\sinh \left(
\kappa \left( r+t/2\right) \right) }+\frac{\pi \sigma ^{2}\cosh ^{2}\left( \frac{%
\kappa t}{2}\right) \widehat{\sum }_{n}a_{n}}{\varepsilon \sinh ^{2}\left( \kappa \left( r+\frac{t}{2%
}\right) \right) }  \label{fm}
\end{eqnarray}

Equation \ref{fm} shows that a swelling of the system ($h=\widehat{\sum }%
_{n}a_{n}<0$) decreases the membrane-membrane interaction whereas
a compression ($h>0$) is unfavorable. Note that we neglected terms
of second
order in the $a_{n}$. As can be seen from Eq. \ref{bend} terms of the form $%
a_{n}^{2}$ lead to a renormalization of the bending constant,
$k_{c}^{\prime }=k_{c}+\left. \delta k\right| _{el}$. It can be
shown that $\left. \delta k\right| _{el}=3\pi \sigma
^{2}/8\varepsilon \kappa ^{3}\approx T/\kappa ^{3}l_{B}\lambda
_{GC}^{2}$ ($\lambda _{GC}=e/2\pi l_{B}\sigma $ is the
Gouy-Chapman length)\cite
{Winter88,Lekker89,Kiometzis89,Fogden90,Pincus90,Higgs90,Duplantier90,Harden92,Andelman95}%
. For a wide range of parameters one has $\left. \delta k\right|
_{el}\ll k_{c}$. In the following we use the bare bending rigidity
$k_{c}$, keeping in mind that it has to be replaced by
$k_{c}^{\prime }$ when $\left. \delta k\right| _{el}$ is
comparable to $k_{c}$.

We estimate now the contribution of the membrane--rod attraction.
Consider the rod at the position $x=0$ and $z=-r-t/2+h$. The rod
is located within an infinite stack of membranes. This can be
accounted for by simply summing {\it twice} over the contributions
of all the membranes that are located {\it above} the rod, i.e.
$f_{MR}=-\left( 2\rho /d\right) \sum\nolimits_{k=0}^{\infty }\Phi
_{M}^{\left( k\right) }\left( 0,-r-t/2+h\right) $. From Eq.
\ref{phim} follows that $\Phi _{M}^{\left( k\right) }\left(
0,z\right) $ is given by (note that $z<0$)
\begin{eqnarray}
&\Phi _{M}^{\left( k\right) }\left( 0,z\right)& \simeq \frac{2\pi \sigma }{%
\varepsilon \kappa }\cosh \left( \frac{\kappa t}{2} \right)
e^{\kappa \left( z-2k\left( r+t/2-h \right) \right) }+\left(
-1\right) ^{k+1} \nonumber \\
&&\frac{2\pi \sigma \kappa }{\varepsilon }\widehat{\sum }_{n}\frac{a_{n}}{%
\kappa _{n}}\cosh \left( \frac{\kappa _{n}t}{2}\right) e^{\kappa
_{n}\left( z-2k\left( r+\frac{t}{2}\right) \right) } \label{potk}
\end{eqnarray}
The contribution of the first term of Eq. \ref{potk} to $F_{MR}$
is of the form:
\begin{eqnarray}
f_{MR}^{\left( 1\right) } &\simeq &-\frac{\sigma \rho q}{\varepsilon \kappa }%
\frac{\cosh \left( \kappa t/2\right) }{\sinh \left( \kappa \left(
r+t/2\right) \right) }-  \nonumber \\ &&\frac{\sigma \rho
q}{\varepsilon }\frac{\coth \left( \kappa \left(
r+\frac{t}{2}\right) \right) }{\sinh \left( \kappa \left( r+\frac{t}{2}\right) \right) }%
\cosh \left( \frac{\kappa t}{2}\right) \widehat{\sum }_{n}a_{n}
\label{fmr1}
\end{eqnarray}
The second term of Eq. \ref{potk} leads to the following
expression:
\begin{equation}
f_{MR}^{\left( 2\right) }=\frac{\sigma \rho q\kappa }{\varepsilon }\widehat{%
\sum }_{n}\frac{\cosh \left( \kappa _{n}t/2\right) }{\kappa
_{n}\cosh \left( \kappa _{n}\left( r+t/2\right) \right) }a_{n}
\label{fmr2}
\end{equation}
The total membrane--DNA contribution $f_{MR}=f_{MR}^{\left(
1\right) }+f_{MR}^{\left( 2\right) }$ favors a compression of the
lamellar stack -- thus constituting a competing mechanism to the
membrane-membrane repulsion.

We are left with the calculation of the interaction energy between
the rods.
We focus here one two important cases. {\it Case 1}: $\kappa t\gg 1$ and $%
\kappa d\ll 1$ (''vertical screening''): In this case the
interaction between rods in different layers is negligible
compared to the rod-rod interaction within the same layer. Then it
is sufficient to sum over the contributions of all rods to the
left and to the right of the given rod :
\begin{equation}
f_{R}=\frac{\rho ^{2}q}{\pi \varepsilon }\sum_{k=1}^{\infty
}K_{0}\left( \kappa dk\right) \simeq \frac{\rho ^{2}q}{\pi \varepsilon }%
\int_{0}^{\infty }dkK_{0}\left( \kappa dk\right) =\frac{\rho
^{2}q^{2}}{4\pi \varepsilon \kappa }  \label{fr2b}
\end{equation}

{\it Case 2}: $\kappa t\ll 1$ and $\kappa d\ll 1$ (weak
screening): In this case all rods contribute to the interaction
energy. After some algebra we arrive at
\begin{equation}
f_{R}\simeq \frac{\rho ^{2}q^{2}}{4\pi \varepsilon \kappa
^{2}\left( r+t/2\right) }+\frac{\rho ^{2}q^{2}}{4\pi \varepsilon
\kappa ^{2}\left( r+t/2\right) ^{2}}\widehat{\sum }_{n}a_{n}
\label{fr3}
\end{equation}
As expected, in both cases the repulsive rod-rod interaction
favors swelling.

\section{Undulations in isoelectric complexes}

\label{sec:2} We consider first the lamellar complex in
equilibrium with a solution of free DNA strands. We ask: What is
the interaxial distance $d$ between the DNA strands in the
lipoplex that minimizes the electrostatic free energy of the
complex? As pointed out by Bruinsma \cite{bruinsma98} counterion
release will control $d$; here, however, we determine the
equilibrium spacing for the case when there is no counterion
release, i.e., we assume the rods being below the Manning
threshold. We also neglect entropic changes due to the adsorption
of free DNA strands into the lipoplex. We will show that, as a
result of the geometry, such a system will equilibrate at the
isoelectric point -- if the electrostatic interaction is
sufficiently long-ranged. In
the following we only account for the contributions independent of the $%
a_{n} $'s and treat the contribution of membrane bending afterwards as a
perturbation.

Let us first consider the case of high ionic strength where $\kappa r\gg 1$
(strong screening). Then the free energy per area is given by
\begin{equation}
f_{tot}\simeq -\frac{4\pi \sigma \rho e^{-\kappa r}}{\varepsilon \kappa d}
\label{ftot}
\end{equation}
i.e., by the membrane-rod attraction, Eq. \ref{fmr1}; other terms are
negligible. It follows that the minimum is at $d\rightarrow 0$. Excluded
volume interaction between the rods will lead to $d=2r$. Clearly, in the
case of strong screening as a result of the short range of the electrostatic
interaction the lipoplex is equilibrated far from the isoelectric point. The
resulting complex is ''overcharged'' by the DNA rods. (A similar situation
occurs for the adsorption of rods on an oppositely charged surface, cf. Ref.
\cite{Nguyen00}).

We discuss next the two cases introduced above. \textit{Case 1}: $\kappa
t\gg 1$ and $\kappa d\ll 1$ (vertical screening): From Eqs. \ref{fmr1} and
\ref{fr2b} we find (up to terms of the order $\kappa d$)
\begin{equation}
f_{tot}\simeq -\frac{2\pi \sigma \rho }{\varepsilon \kappa d}+\frac{\pi \rho
^{2}}{\varepsilon \kappa d^{2}}  \label{ftot2}
\end{equation}
$f_{tot}$ is minimized for $d=d_{iso}=\rho /\sigma $ which corresponds to
the isoelectric point of the complex, i.e., the point at which the charges
of the cationic lipids and of the DNA are exactly balanced. \textit{Case} 2:
$\kappa t\ll 1$ and $\kappa d\ll 1$ (weak screening): From Eqs. \ref{fmr1}
and \ref{fr3} follows
\begin{equation}
f_{tot}\simeq -\frac{2\pi \sigma \rho }{\varepsilon \kappa ^{2}\left(
r+t/2\right) d}+\frac{\pi \rho ^{2}}{\varepsilon \kappa ^{2}\left(
r+t/2\right) d^{2}}  \label{ftot3}
\end{equation}
Again the free energy is minimized at the isoelectric interhelical spacing $%
d=d_{iso}=\rho /\sigma $.

Thus in the limiting case $\kappa \rightarrow 0$ (no salt, no screening) the
lipoplex is forced to be at the isoelectric point. This is an artefact of
our model which does not account for counterions that would be present in
lipoplexes with an excess of DNA molecules or cationic lipids. This holds
for both cases, for Case 1 corresponding effectively to decoupled
two-dimensional layers and for Case 2 which is truly three-dimensional.
Therefore our theory is only applicable to isoelectric lipoplexes where all
counterions are expected to be released.

We consider now the undulations occuring in lipoplexes in general and then
focus again on the isoelectric point. The change of the total electrostatic
free energy as a function of the deformation follows from the Eqs. \ref{fm},
\ref{fmr1}--\ref{fr3}:
\begin{equation}
\Delta f_{el}\simeq \left\{
\begin{array}{ll}
\left( \frac{\pi \sigma ^{2}}{\varepsilon }-\frac{\sigma \rho
q}{\varepsilon }\right) \widehat{\sum }_{n}a_{n} &
\mbox{for}\,\,\kappa t\gg 1,\kappa d\ll 1
\\
\left( \frac{\pi \sigma ^{2}}{\varepsilon }-\frac{\sigma \rho
q}{\varepsilon
}+\frac{\rho ^{2}q^{2}}{4\pi \varepsilon }\right) \frac{\widehat{\sum }%
_{n}a_{n}}{\kappa ^{2}\left( r+\frac{t}{2}\right) ^{2}} & \mbox{for}%
\,\,\kappa t\ll 1,\kappa d\ll 1
\end{array}
\right.   \label{fel3}
\end{equation}
The change of the total free energy due to bending is given by the sum of
the electrical contribution $\Delta f_{el}$ and the bending energy $f_{bend}$%
, Eq. \ref{bend}. Minimizing $f_{tot}$ with respect to the amplitudes $a_{n}$
leads to $a_{n}\simeq A/n^{4}$ with
\begin{equation}
A\simeq \left\{
\begin{array}{ll}
\frac{\sigma \rho q-\pi \sigma ^{2}}{\varepsilon k_{c}q^{4}} & \mbox{for}%
\,\,\kappa t\gg 1,\;\kappa d\ll 1 \\
\frac{\sigma \rho q-\pi \sigma ^{2}-\rho ^{2}q^{2}/4\pi }{\varepsilon
k_{c}q^{4}\kappa ^{2}\left( r+t/2\right) ^{2}} & \mbox{for}\,\,\kappa t\ll
1,\;\kappa d\ll 1
\end{array}
\right.  \label{a}
\end{equation}
Thus the deformation modes decrease rapidly with increasing $n$. Now we are
in the position to calculate the deformation of the membranes. Inserting $%
a_{n}=A/n^{4}$ into Eq. \ref{prof} we find the following deformation profile
\begin{equation}
u\left( x\right) =A\widehat{\sum }_{n}\frac{\cos \left( nqx\right) }{n^{4}}%
=A\left( \frac{\pi ^{4}}{90}-\frac{\pi ^{2}x^{2}}{12}+\frac{\pi x^{3}}{12}-%
\frac{x^{4}}{48}\right)  \label{prof2}
\end{equation}
where the polynomial expression is valid for $0\leq qx\leq \pi $ (the
continuation outside this interval follows from the symmetry of the
configuration). A good approximation for all values of $x$ (relative error
smaller than $1.5\%$) is given by $u\left( x\right) =A\cos \left( qx\right) $%
.

The coefficient $A$ at the isoelectric point of the complex $2\pi \sigma
/q=\rho $ is given by
\begin{equation}
A\simeq \left\{
\begin{array}{ll}
\frac{\pi \sigma ^{2}}{\varepsilon k_{c}q^{4}} & \mbox{for}\,\,\kappa t\gg
1,\;\kappa d\ll 1 \\
0 & \mbox{for}\,\,\kappa t\ll 1,\;\kappa d\ll 1
\end{array}
\right.  \label{a2}
\end{equation}
In the case of vertical screening we find a positive (and $\kappa $%
-independent) value of $A$ and thus a positive value of $h$, $h=\pi ^{4}A/90$%
, corresponding to a compression of the isoelectric lamellar stack.
Interestingly, in the case of weak screening the undulations disappear. In
fact, as long as the vertical screening is operative the membrane--membrane
repulsion is smaller than the membrane-rod attraction in the isoelectric
lipoplex and therefore we find a compression of the lamellar stack. For weak
screening, the rod-rod repulsion between different layers cancels this net
attraction, cf. Eq. \ref{fel3}. Let us consider typical values for $\rho $, $%
\sigma $, $\varepsilon $ and $k_{c}$, say $\rho =e/1.7$\AA\ (DNA), $\sigma
=e/100$\AA $^{2}$, $\varepsilon =80$ (water) and $k_{c}=20k_{B}T$. For these
values we find (in the case of vertical screening) $h\approx 1$\AA , i.e.,
the undulations are rather small.

Finally, we estimate how the undulations of the isoelectric complex disturb
the interaxial spacing between the rods and in turn move the complex away
from its isoelectric point. We consider the case of vertical screening (Case
1, $\kappa t\gg 1$, $\kappa d\ll 1$). In this case the amplitudes of the
undulations depend strongly on the interhelical distance, namely $A\sim
d^{4} $, cf. Eq. \ref{a2}. Inserting $h=\widehat{\sum }_{n}a_{n}=\pi ^{4}A/90
$ into Eq. \ref{fel3} ($\kappa t\gg 1$, $\kappa d\ll 1$) we find two
correction terms to Eq. \ref{ftot2}, namely $\pi ^{2}\sigma ^{4}d^{4}/\left(
1440\varepsilon ^{2}k_{c}\right) $ from the membrane-membrane repulsion and $%
-\pi ^{2}\sigma ^{3}\rho d^{3}/\left( 720\varepsilon
^{2}k_{c}\right) $ from the membrane-rod attraction. Evidently,
the mem-brane-membrane repulsion favors smaller values of $d$ that
lead to smaller undulations, whereas the membrane-rod interaction
is enhanced for larger undulations, i.e., larger values of $d$ are
favorable. At the isoelectric point $d=\rho /\sigma $ the
correction term from the membrane-rod interaction exceeds the
other term and
as a result the interhelical distance is slightly increased, $%
d=d_{iso}+\Delta d$, with
\begin{equation}
\Delta d=\frac{\pi }{720}\frac{\kappa \rho ^{5}}{\varepsilon k_{c}\sigma ^{3}%
}  \label{dd}
\end{equation}
In that sense the undulations lead to an effective repulsion between the DNA
strands proportional to $d^{3}$ (as long as one is close enough to the
isoelectric point). Undulations are one of several mechanisms that might be
responsible for the increase of the interhelical spacing with increasing
salt concentration as it is observed experimentally (cf. Ref. \cite
{Koltover99} for details).

\section{Discussion}

\label{sec:3} The main idea of the preceding analysis is to give a
simple estimate of the role of the electrostatic interactions in a
lamellar lipoplex. In many instances some of the underlying
assumptions are not fulfilled. But even in this case our model
might give an idea about what the contributions of the
electrostatics to the overall conformation might be.

One severe approximation is the assumption of a transparent membrane. The
lipid bilayer represents a low dielectric slab that -- depending on its
thickness -- might screen most of the electrical field so that charges (say
of phosphate group on a DNA molecule) are not ''seen'' on the other side of
the membrane. As a rule of thumb, a membrane that is much thinner than the
distance $D$ of a charge from the membrane might be considered to be
transparent for this charge whereas a thicker membrane (of thickness $t>D$)
is opaque and can be approximated by an infinitely thick slab. In that case
the effect of the low dielectric lipid can be accounted for by the use of an
appropriate image charge -- which is a simple task for a flat membrane but
difficult to handle for an undulating one. The two cases $t<D$ and $t>D$ are
elaborated in some detail in Footnote 2 in Ref. \cite{Schiessel98}. The
typical thickness of a lipid membrane is $24$\AA\ which is of the order of
the diameter of the DNA rod ($20$\AA ), i.e., one is in the crossover regime
between the two cases. In any case, the presence of the low-dielectric lipid
will lead to a modification of the simple situation discussed in this paper.
It should be expected that the partial confinement of electrical field lines
emanating from the DNA rods by the neighboring membranes favors a swelling
of the lamellar stack.

Another feature not considered in this study is the demixing of neutral and
charged lipids within the bilayers. There might be at least three effects. (%
\textit{i}) The electrostatic attraction drives the cationic lipids towards
the DNA rods, resulting in a depletion of charged lipids in the membrane
parts in between two neighboring rods in the same layer. This reduces the
membrane-membrane repulsion between neighboring membranes resulting in an
increase of the compression of the lipoplex. This effect should be important
if the average mole fraction of cationic lipids in the bilayers is low. (%
\textit{ii}) For the opposite limit of highly charged membranes their
surface charge density $\sigma $ might exceed the surface charge density $%
\rho /2\pi r$ of the rods. In this case an enhancement of neutral lipids
close to the DNA is expected that allows for a better matching of the two
charge densities. These two effects were indeed observed in the numerical
study by Harries \textit{et al.} \cite{Harries98} (cf. Fig. 7 in that
paper). (\textit{iii}) Finally, membrane undulations may also affect the
charge densities on each side of the bilayer and vice versa. A depletion of
cationic lipids on one side of the membrane will be accompanied by an
enhancement on the other side due to the symmetry of the arrangement of DNA
rods, cf. Fig. 1. This might lead to a spontaneous curvature of the bilayer
of either sign which in turn affects the membrane undulations discussed in
our study. It is clear that the competition of these three effects can lead
to a rather complex behavior of the charge density profile of the lipids
along the $X$-direction. In our model we do not account for these effects.
We expect that our approach has to be modified especially when a large
fraction of the lipids is neutral. In that case the description of local,
highly charged pinches \cite{Schiessel98} might be more appropriate (cf. the
discussion of the polypeptide lipoplex \cite{Subu00} given in the
Introduction of our paper).

Recently it has been possible to non-ambiguously detect undulations in a
lipoplex\cite{Raedlerpp}. From a careful analysis of the data on the
lipoplex presented in Ref. \cite{Artzner98} if was possible to construct an
electron density map revealing its high resolution structure. The structure
shows undulations with an amplitude of a few Angstrom leading to a
compression of the lamellar structure as depicted in Fig. 2a. Furthermore,
the amplitudes show a sharp increase for larger interhelical spacings. Both
observations are in qualitative and semi-quantitative agreement with the
results of our model calculation, cf. Eq. \ref{a2}. It is worth noting that
the lipids of these lipoplexes form a lipid-gel phase \cite{Artzner98}. In
this phase the lipid bilayers show a high compressibility allowing the
DNA-induced deformations to cross nearly unperturbed through the bilayer (as
implicitely assumed in our model). The induced undulations might be the
prevailing mechanism for the interlayer coupling that leads to the
rectangular columnar superlattice of the DNA strands observed for this class
of lipoplexes.

\textit{Acknowledgement}: We wish to thank J. O. R\"{a}dler for
sharing experimental results prior to publication. We would like
to acknowledge useful conversations with A. Ben-Shaul and S. May.


\begin{thebibliography}{99}
\bibitem{Felgner97}  P.L. Felgner, Sci. Am.\textit{\ }\textbf{276}, 102
(1997).

\bibitem{Crystal95}  R.G. Crystal, Science \textbf{270}, 404 (1995).

\bibitem{Sternberg94}  B. Sternberg, F. Sorgi, L. Huang, FEBS Lett.\textit{\
}\textbf{356}, 361 (1994).

\bibitem{lasic97}  D.D. Lasic, H. Strey, M.C.A. Stuart, R. Podgornik, P.M.
Frederik, J. Am. Chem. Soc. \textbf{119}, 832 (1997).

\bibitem{Raedler97}  J.O. R\"{a}dler, I. Koltover, T. Salditt, C.R. Safinya,
Science\textit{\ }\textbf{275}, 810 (1997).

\bibitem{fang97}  Y. Fang, J. Yang, J. Phys. Chem. B\textit{\ }\textbf{101},
441 (1997); J. Phys. Chem. B\textit{\ }\textbf{101}, 3453 (1997).

\bibitem{salditt97}  T. Salditt, I. Koltover, J.O. R\"{a}dler, C.R. Safinya,
Phys. Rev. Lett.\textit{\ }\textbf{79}, 2582 (1997); Phys. Rev. E\textit{\ }%
\textbf{58}, 889 (1998).

\bibitem{koltover98}  I. Koltover, T. Salditt, J.O. R\"{a}dler, C.R.
Safinya, Science\textit{\ }\textbf{281},78 (1998).

\bibitem{Artzner98}  F. Artzner, R. Zantl, G. Rapp, J.O. R\"{a}dler, Phys.
Rev. Lett.\textit{\ }\textbf{81}, 5015 (1998).

\bibitem{Zantl98}  R. Zantl, F. Artzner, G. Rapp, J.O. R\"{a}dler, Europhys.
Lett.\textit{\ }\textbf{45}, 90 (1999).

\bibitem{huebner99}  S. Huebner, B.J. Battersby, R. Grimm, G. Cevc, Biophys.
J.\textit{\ }\textbf{76}, 3158 (1999).

\bibitem{Koltover99}  I. Koltover, T. Salditt, C.R. Safinya, Biophys. J.%
\textit{\ }\textbf{77}, 915 (1999).

\bibitem{dan96}  N. Dan, Biophys. J. \textbf{71}, 1267 (1996).

\bibitem{dan97}  N. Dan, Biophys. J.\textit{\ }\textbf{73}, 1842 (1997).

\bibitem{bruinsma98b}  R. Bruinsma, J. Mashl, Europhys. Lett.\textit{\ }%
\textbf{41}, 165 (1998).

\bibitem{May97}  S. May, A. Ben-Shaul, Biophys. J.\textit{\ }\textbf{73},
2427 (1997).

\bibitem{dan98}  N. Dan, BBA-Biomembranes \textbf{1369}, 34 (1998).

\bibitem{bruinsma98}  R. Bruinsma, Eur. Phys. J. B\textit{\ }\textbf{4}, 75
(1998).

\bibitem{golubovic98}  L. Golubovi\'{c}, M. Golubovi\'{c}, Phys. Rev. Lett.%
\textit{\ }\textbf{80}, 4341 (1998).

\bibitem{ohern98b}  C.S. O'Hern, T.C. Lubensky, Phys. Rev. Lett.\textit{\ }%
\textbf{80}, 4345 (1998).

\bibitem{Harries98}  D. Harries, S. May, W.M. Gelbart, A. Ben-Shaul,
Biophys. J.\textit{\ }\textbf{75}, 159 (1998).

\bibitem{Schiessel98}  H. Schiessel, Eur. Phys. J. B\textit{\ }\textbf{6},
373 (1998).

\bibitem{menes98}  R. Menes, P. Pincus, R. Pittman, N. Dan, Europhys. Lett.%
\textit{\ }\textbf{44}, 393 (1998).

\bibitem{ohern98}  C.S. O'Hern, T.C. Lubensky, Phys. Rev. E\textit{\ }%
\textbf{58}, 5948 (1998).

\bibitem{ohern99}  C.S. O'Hern, T.C. Lubensky, J. Toner J., Phys. Rev. Lett.%
\textit{\ }\textbf{83}, 2745 (1999).

\bibitem{may00}  S. May, D. Harries, A. Ben-Shaul, Biophys. J.\textit{\ }%
\textbf{78}, 1681 (2000).

\bibitem{menes00}  R. Menes, N. Gr\o nbech-Jensen, P.A. Pincus, Eur. Phys.
J. E \textbf{1}, 345 (2000).

\bibitem{park99}  S. Park, R.F. Bruinsma, W.M. Gelbart, Europhys. Lett.%
\textit{\ }\textbf{46}, 454 (1999).

\bibitem{Schiessel00}  H. Schiessel, R. Bruinsma, W. M. Gelbart, preprint

\bibitem{Subu00}  G. Subramanian, R.P. Hjelm, T.J. Deming, G.S. Smith, Y.
Li, C.R. Safinya, J. Am. Chem. Soc.\textit{\ }\textbf{122}, 26 (2000).

\bibitem{Verwey}  E.J.W. Verwey, J.Th.G. Overbeck, \textit{Theory of the
Stability of Lyophobic Colloids} (Elsevier, Amsterdam,1948).\noindent
\noindent

\bibitem{Goldstein90}  R.E. Goldstein, A.I. Pesci, V. Romero-Rochin, Phys.
Rev. A \textbf{41}, 5504 (1990).\noindent

\bibitem{Safran}  S.A. Safran, \textit{Thermodynamics of Surfaces,
Interfaces, and Membranes} (Addison-Wesley, Reading, MA, 1994).\noindent

\bibitem{Winter88}  M. Winterhalter, W. Helfrich, J. Phys. Chem. \textbf{92}%
\textit{, 6865 (}1988).\noindent

\bibitem{Lekker89}  H.N.W. Lekkerkerker, Physica A \textbf{159}, 319 (1989).

\bibitem{Kiometzis89}  M. Kiometzis, H. Kleinert, Phys. Lett. A \textbf{140}%
, 520 (1989).

\bibitem{Fogden90}  A. Fogden, D.J. Mitchell, B.W. Ninham, Langmuir \textbf{6%
}, 159 (1990).\noindent

\bibitem{Pincus90}  P. Pincus, J.-F. Joanny, D. Andelman, Europhys. Lett.
\textbf{11}, 763 (1990).

\bibitem{Higgs90}  P.G. Higgs, J.-F. Joanny, J. Phys. France \textbf{51},
2307 (1990).

\bibitem{Duplantier90}  B. Duplantier, Physica A \textbf{168}, 179 (1990).

\bibitem{Harden92}  J.L. Harden, C. Marques, J.-F. Joanny, D. Andelman,
Langmuir \textbf{8}, 1170 (1992).

\bibitem{Andelman95}  D. Andelman in: \textit{Structure and Dynamics of
Membranes.} Eds. R. Lipowsky and E. Sackmann (North-Holland, 1995).

\bibitem{Nguyen00}  T.T. Nguyen., A.Yu. Grosberg, B.I. Shklovskii, J. Chem.
Phys.\textit{\ }\textbf{113}, 1110 (2000).

\bibitem{Raedlerpp}  J.O. R\"{a}dler, unpublished results
\end{thebibliography}
\end{document}